\documentclass[10pt, conference, letterpaper]{IEEEtran}
\usepackage{booktabs} 
\usepackage{url}
\usepackage{rotating}
\usepackage{graphicx}
\graphicspath{{figures/}}

\usepackage{float,array}
\usepackage{mwe,tikz}
\usepackage{amsmath}

\usepackage{amsthm}
\usepackage{float}
\usepackage{enumitem}

\usepackage{amssymb}
\usepackage{algpseudocode}
\usepackage{algorithm}
\usepackage{epstopdf}

\usepackage{multirow}
\usepackage[font={small}]{caption}

\usepackage[font={small}]{subfig}
\usepackage{balance}
\usepackage{color}

\usepackage{bbm}

\IEEEoverridecommandlockouts
\begin{document}
\title{Jupiter: A Networked Computing Architecture}

\author{\IEEEauthorblockN{Pradipta Ghosh, 
        Quynh Nguyen,
        Pranav K Sakulkar,
        Aleksandra Knezevic,
        Jason A. Tran,
        Jiatong Wang
    }\\
    \IEEEauthorblockN{
        Zhifeng Lin,
        Bhaskar Krishnamachari,
        Murali Annavaram,
        and Salman Avestimehr
    }
    \thanks{All authors are with the Ming Hsieh Department of Electrical Engineering, University of Southern California, Los Angeles, CA, USA.}
    \thanks{Corresponding Email: pradiptg@usc.edu}
    \thanks{This material is based upon work supported by Defense Advanced Research Projects Agency (DARPA) under Contract No. HR001117C0053. Any views, opinions, and/or findings expressed are those of the author(s) and should not be interpreted as representing the official views or policies of the Department of Defense or the U.S. Government.}
}

\maketitle

\begin{abstract}
In the era of Internet of Things, there is an increasing demand for networked computing to support the requirements of the time-constrained, compute-intensive distributed applications such as multi-camera video processing and data fusion for security. We present Jupiter, an open source networked computing system that inputs a Directed Acyclic Graph (DAG)-based computational task graph to efficiently distribute the tasks among a set of networked compute nodes regardless of their geographical separations and orchestrates the execution of the DAG thereafter. This Kubernetes container-orchestration-based system supports both centralized and decentralized scheduling algorithms for optimally mapping the tasks based on information from a range of profilers: network profilers, resource profilers, and execution time profilers. While centralized scheduling algorithms with global knowledge have been popular among the grid/cloud computing community, we argue that a distributed scheduling approach is better suited for networked computing due to lower communication and computation overhead in the face of network dynamics. To this end, we propose and implement a new class of distributed scheduling algorithms called WAVE on the Jupiter system. We present a set of real world experiments on two separate testbeds - one a world-wide network of 90 cloud computers across 8 cities and the other a cluster of 30 Raspberry pi nodes, over a simple networked computing application called Distributed Network Anomaly Detector (DNAD). We show that despite using more localized knowledge, a distributed WAVE greedy algorithm can achieve similar performance as a classical centralized scheduling algorithm called Heterogeneous Earliest Finish Time (HEFT), suitably enhanced for the Jupiter system. 
\end{abstract}

\section{Introduction}

With the miniaturization of hardware in the era of Internet of Things (IoT), the presence of economical low-compute-power edge devices such as cell phones, car dashboard, and drones have become ubiquitous near end users.
This has opened up the domain of edge or fog computing~\cite{mao2017survey} that focuses on exploiting all the devices near end users to comply with the skyrocketing demand for computationally intensive applications such as image processing and voice recognition towards autonomy and personalized assistance.
Interestingly, a significant subset of these cutting-edge time-constrained, compute-intensive distributed applications rely on an orderly processing of the streaming data generated from a set of devices that maybe geographically dispersed. 
This brings us to the newly emerging field of \emph{Networked Computing} or \emph{Dispersed Computing} that focuses on a joint optimization of computation and communication costs to distribute the execution of a Directed Acyclic Graph (DAG) based task graph among a network of compute nodes that may be geographically distributed. 
Networked Computing can be thought of as a mixed architecture between Edge Computing and Cloud Computing where the network of compute nodes might contain either or both edge processors and cloud-based processors.
This new field of Networked computing calls for a distributed system that can optimally leverage the available compute resources in a network of compute nodes while accounting for any network delays that may impact the timely processing of  data. 

In this paper, we present Jupiter, an open-source system for networked computing that contains the necessary tools to efficiently map the tasks from a task-DAG into a set of geographically distributed networked compute processors (NCPs) with the main focus on \emph{`Makespan'} minimization.
We define the \emph{`Makespan'} of the DAG as the time required to generate an output via executing the entire task DAG on one set of input files or input chuck of data.
Jupiter also administers the actual processing of the tasks along with efficient data transfer between them. 
Jupiter relies on the reputed open-source container-orchestrator tool from Google called Kubernetes~\cite{hightower2017kubernetes} for implementation of the three main components: (1) Profilers that gather statistics about the network condition, resource availability in the NCPs, and execution time of the tasks on the NCPs, (2) Task Mapper that leverages the information available from three different types of profiler modules to optimally schedule or map the tasks into the NCP nodes to minimize the Makespan of the DAG, and (3) CIRCE that boots-up the tasks according to the task mapping and administers the task executions and data transfers.
For task to NCP mapping, Jupiter has plug-n-play type provisions for both centralized and decentralized mapping or scheduling algorithms.
While centralized mappers such as the Heterogeneous Earliest Finish Time (HEFT)~\cite{topcuoglu2002performance} are proved to be promising and efficient for cloud/grid computing, we argue that a distributed scheduler with comparable performance is more appropriate for networked computing systems. 
To this end, we propose the WAVE framework that is a novel class of decentralized task mapping algorithms which we demonstrate to have similar empirical statistics as the HEFT algorithm. 

To test the performance of the Jupiter system for varying network and resource conditions, we perform a wide range of experiments on two fairly large testbeds: (1) a 90 node testbed based on the DigitalOcean cloud platform where we handpicked the servers from a set of 8 geographically distributed cluster locations, and (2) a 30 node in-house Raspberry Pi3 (RPI3) cluster connected via a Cisco switch to control the network characteristics. 
For the experiments, we have also implemented a sample networked computing application called the Distributed Network Anomaly Detection (DNAD) which is focused on real-time defence against Distributed Denial of Service (DDoS) attacks in a network. 
Our experiments show that for highly resource-constrained devices, such as the Raspberry Pi3, the original HEFT performs quite poorly.
This led us to design a modified version of HEFT that has better performance in such resource-constrained devices. 

In summary, our contributions in this paper can be listed as follows:
\begin{itemize}
    \item We propose a novel open-source networked computing system called the Jupiter that supports proper profiling of the resources, efficient centralized/decentralized task-to-compute node mapping for an application-DAG, and administered execution of the application-DAG. 
    \item We propose a new class of distributed local-information-based scheduling/mapping algorithms called WAVE that has similar performance to a well-known centralized, globally informed heuristic called the HEFT.
    \item We formulate a new application for networked computing called the Distributed Network Anomaly Detection (DNAD) which is focused on real-time defence against Distributed Denial of Service (DDoS) attacks in a network.
    \item Based on a range of experiments, we discover some shortcomings of the HEFT like algorithms on the Jupiter system and, thereafter, propose some constructive modifications. 
    
\end{itemize}
\begin{figure}[!ht]
    \centering
    \includegraphics[width=\linewidth]{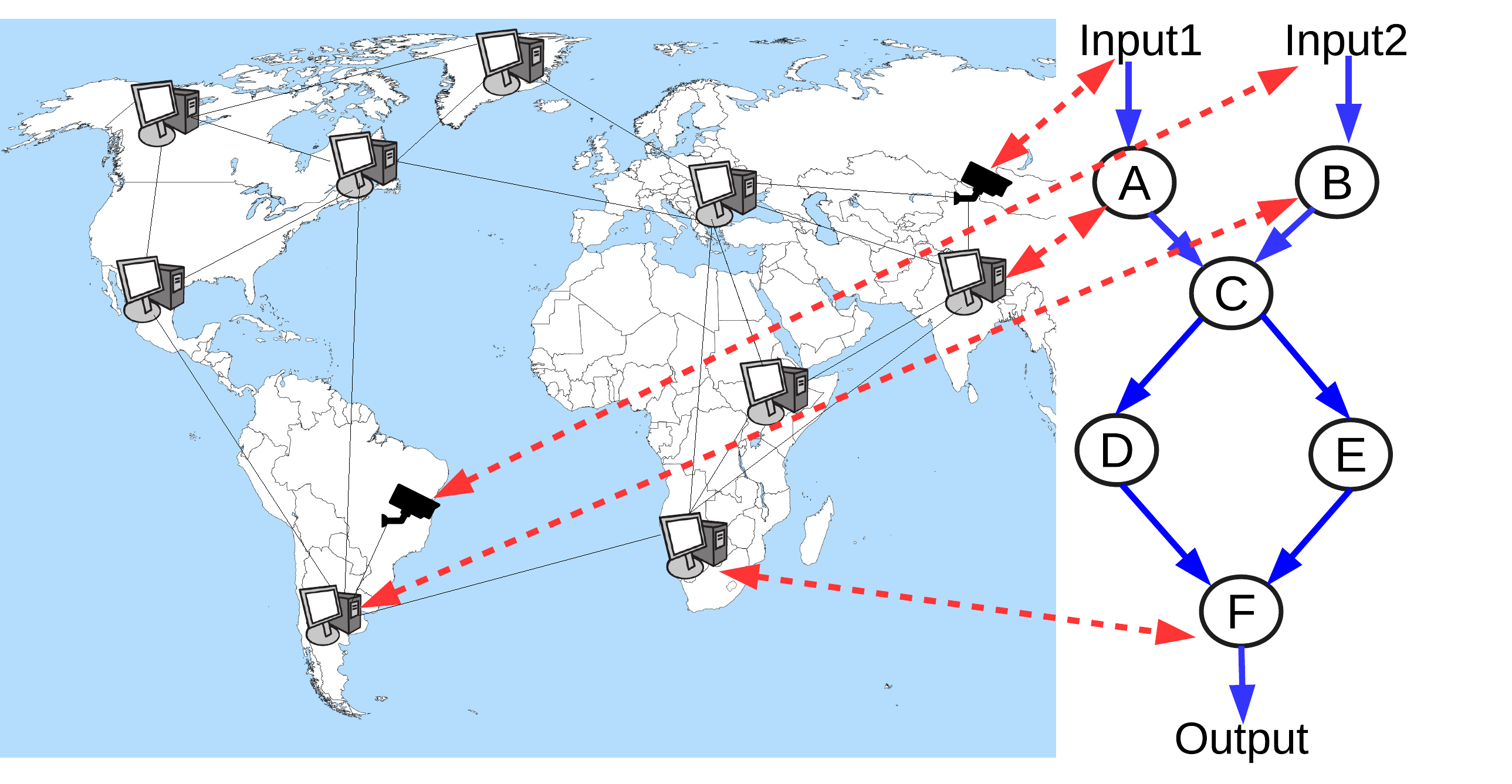}
    \caption{Illustration of the DAG based Networked Computing problem. The black lines denote communication links, the red lines denote the mapping, and the blue lines denote data flows in the DAG. }
    \label{fig:problem}
\end{figure}

\section{Problem Description}
First of all, let us assume that we have a network of $N$ heterogeneous networked compute processors (NCPs) that are geographically distributed across the world.
Therefore, the end-to-end latency between these NCPs are statistically different as well as dynamic.
Let's say, we are interested in deploying an application-DAG that consists of $T$ tasks where the input sources are distributed across the world.  
Now, the goal is to properly map the tasks from the application DAG to the NCPs such that the Makespan of the application-DAG is minimized. 
The Makespan in this context would depend on both the compute powers of the NCPs chosen as well as the delays on the network paths between the NCPs.
For an illustration, refer to Fig.~\ref{fig:problem} where the application DAG consists of 6 tasks with two geographically separated input sources. 
Now, the goal is to optimally map the tasks on the geographically distributed NCPs such that the output can be made available the fastest.

\subsection{The DNAD Application}
\begin{figure}[!ht]
    \centering
    \includegraphics[width = 0.8\linewidth]{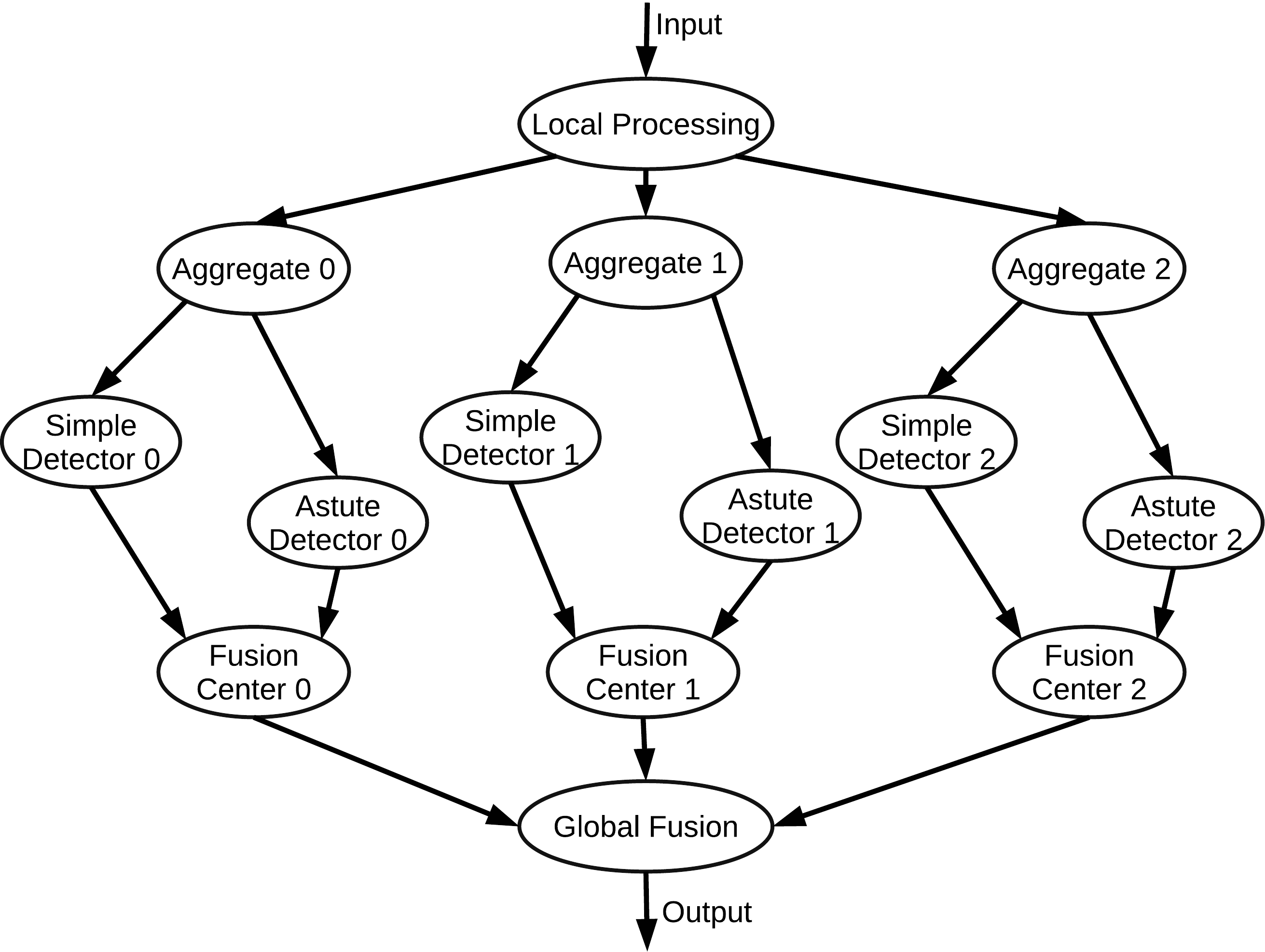}
    \caption{The DAG for the DNAD Application}
    \label{fig:dnad}
\end{figure}

We present a sample application for Networked Computing called the Distributed Networked Anomaly Detector (DNAD).
The main goal of this application is to use a network of computation capable routers to detect Distributed Denial of Service (DDoS) attacks. 
We assume that a set of border routers of a distributed administrative network monitors the incoming traffic to periodically generate traffic statistics which can be employed to detect the potentially anomalous IP addresses and to take affirmative actions against such IP addresses. 
One can easily have a powerful central server to collect the data and process it. 
However, for a geographically dispersed network, the communication delay between different border routers and the cloud is significant which makes the process non-realtime as well as costly due to the usage of cloud resources.
The networked computing based solution can come to the rescue by using all the compute resources available on the network, such as the processors on the routers, towards realtime processing of the traffic statistics and realtime protection of the network against such DDoS attacks.
The task DAG for the DNAD is presented in Fig.~\ref{fig:dnad}. Next, we briefly explain each of the tasks. 

\textbf{Local Processing:}
The local processing task node is mainly the observation point of the network which monitors the incoming traffic and collects the statistics by hashing the traffic statistics using the IP addresses. Next, it splits the input data stream into three independent streams based on the evaluated hash values. Each stream is further sent to one of the three child tasks i.e., aggregation points (illustrated in Fig.~\ref{fig:dnad}). 

\textbf{Aggregation Point:}
In our DNAD implementation, we have three aggregate task nodes that collect statistics with the similar hash value from the observation points i.e, local processing units. In this example, we have only one observation point for simplicity. Ideally, the number of aggregation nodes can be any positive number. This number is mainly used for load balancing the detection process. The output of each aggregate node is sent to a range of detector node.

\textbf{Anomaly Detectors:}
Any network anomaly detector is bound to have false alarms and missed detections. Using several detectors in parallel and combining their outputs together is warranted to provide more robustness to the detection performance. We use two different detectors on each data-stream: (1) Simple detector, which simply applies a threshold on the traffic generated from each IP in order to filter out the anomaly and (2) Astute detector, which is an implementation of the Astute anomaly detector~\cite{silveira2011astute}. 
One can easily extend this framework to use more than 2 types of detectors. 

\textbf{Fusion Center:}
Each fusion center aggregates and combines the output of the respective anomaly detectors to output a unified list of anomalous IP addresses for that particular data stream. 

\textbf{Global Fusion:} 
The global fusion unit combines the outputs of the fusion centers to output a file with the final inference about the observed traffic.


\section{The Jupiter Architecture}
\label{sec:jupiter}

\begin{figure}[!ht]
    \centering
    \includegraphics[width=0.8\linewidth]{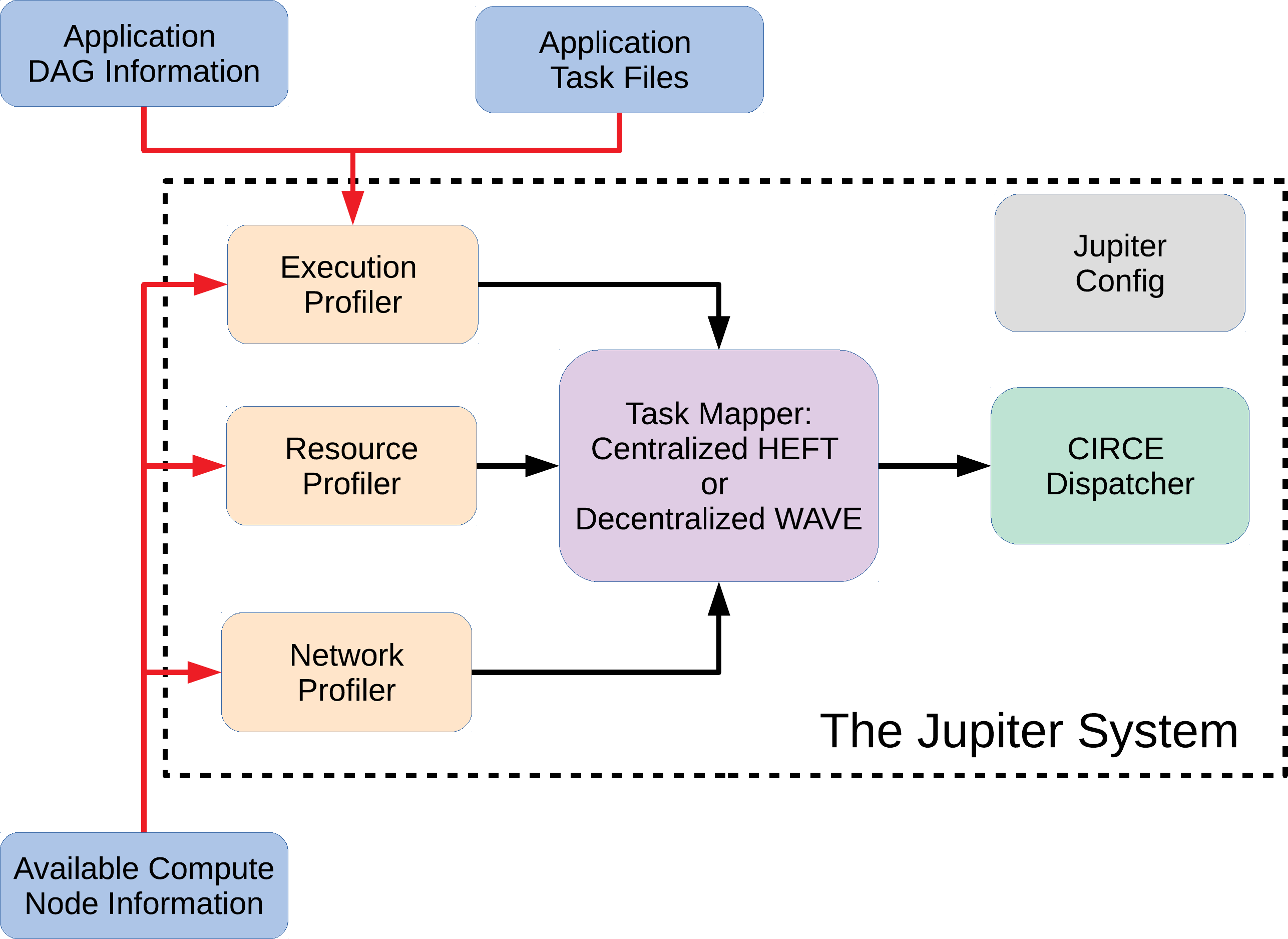}
    \caption{The Jupiter Architecture}
    \label{fig:jupiter}
\end{figure}

Jupiter is a networked computing system to automate the mapping of application DAG to an arbitrary network under both centralized and decentralized settings. 
The Jupiter system consists of three main modules: \emph{Profiler, Task Mapper, and the CIRCE dispatcher}.
The inputs to the Jupiter consist of the Directed Acyclic Task Graph (DAG) information, the task files, and the information (such as IP or node name) about available compute nodes.
The profiler module of the Jupiter consists of three different types of profilers: (1) Network Profiler that maintains statistics about the bandwidth and end-to-end delay between the available NCPs, (2) Resource Profiler that profiles the resource availability of each NCP in terms of CPU and Memory availability, and (3) Execution profiler that profiles the execution time of each task of the DAG in each of the available NCPs. 
The information from the profilers and the input files are fed to the task mapper module which outputs a mapping of the DAG tasks into the available NCPs based on the mapping algorithm used.
Next, the generated task-to-NCP-mapping is used by the CIRCE dispatcher to dispatch the tasks on respective NCPs, monitor the input-output of each task to administer the respective task execution, and transfer the data/files between consecutive tasks of the DAG.
In the current Jupiter system, we have provision for two different classes of task mappers: (1) Centralized HEFT and (2) Decentralized WAVE.
A Jupiter configuration file is used for choosing between these different options of task mappers as well as setting a range of parameters to customize for application-specific requirements.
In the Jupiter architecture, we assume that there exists at least one NCP in the network that can act as an administrative node to the network which we refer to as the ``Home NCP".
The Home NCP can be any randomly selected NCP of the network if no such distinct administrative NCP exists.
In Fig.~\ref{fig:jupiter}, we illustrate the architecture of the proposed Jupiter system along with the data flow between different modules. 

\subsection{Profilers}
\begin{figure}[!ht]
    \centering
    \includegraphics[width=0.9\linewidth]{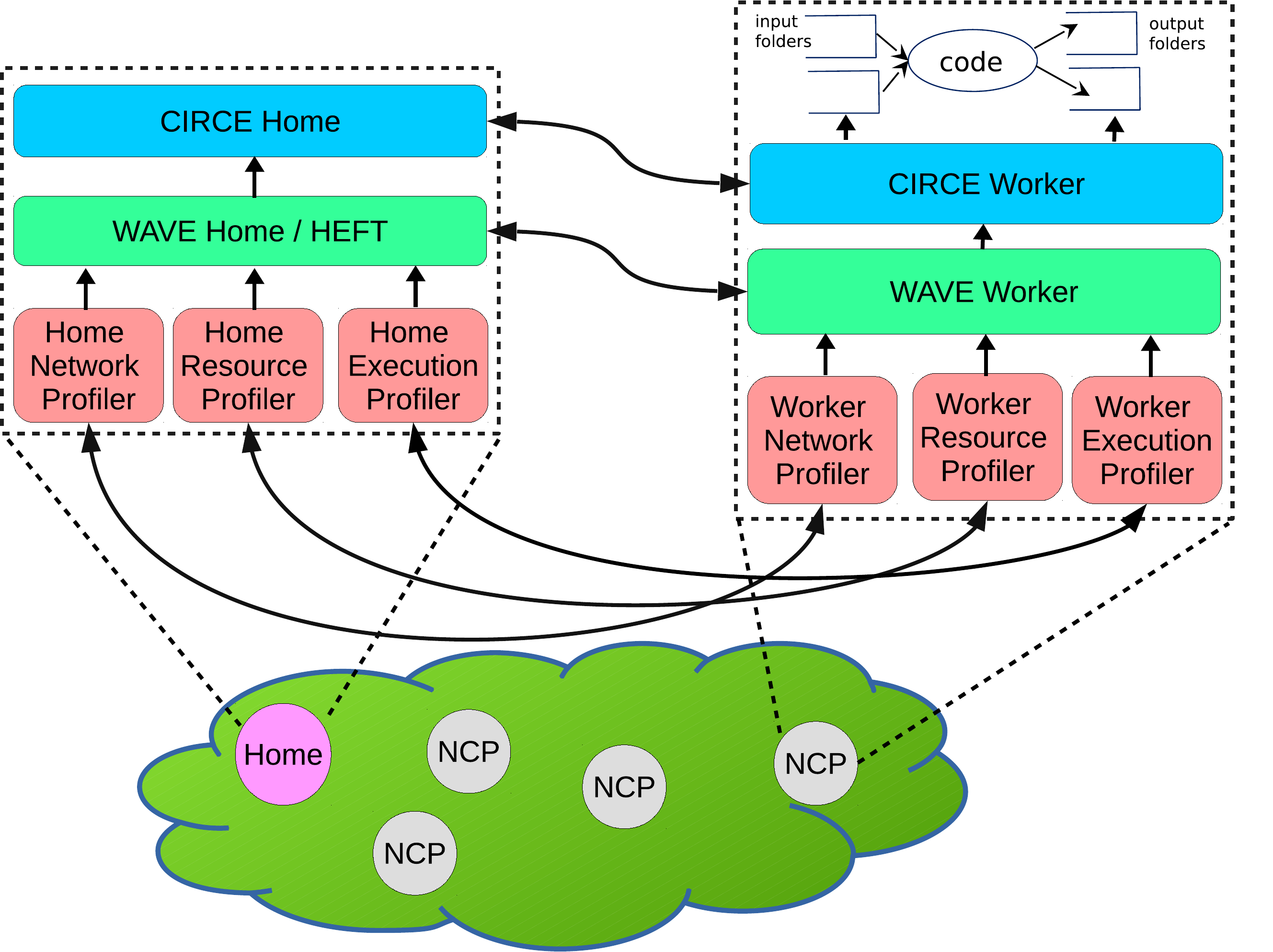}
    \caption{Illustration of the Jupiter deployed system. There exists a Home NCP that runs all the Home Profilers, the Home Task Mapper, and Home CIRCE. The Home CIRCE is used purely for experimental purpose. Similarly, all the other NCPs in the networks run the Worker parts of the Profilers, WAVE (if used), and CIRCE. }
    \label{fig:jupiternetwork}
\end{figure}

To provide support for a broad range of mapping algorithms, Jupiter's profiler module consists of three different types of profilers: Network, Resource, and Execution Profiler. A key architectural component of all three types of profilers is that there exists one Home profiler which runs on the Home NCP, while rest of the NCPs run a copy of the Worker Profiler. 
This is illustrated in Fig.~\ref{fig:jupiternetwork}.
The Home profiler acts as a master to initiate and orchestrate the profiling process while keeping track of available NCP information.
The Worker profilers perform the actual profiling job on each NCP which we discuss in the following.

\subsubsection{Network Profiler}
A major component in the Makespan of a DAG-based application is the file transfer latency between consecutive tasks in the task graph. 
Thus, the end-to-end delay between two NCP nodes is an important parameter for task mapping.
The network profiler in the Jupiter system provides that information by having a network profiling job run on each node, which we also refer to as the Worker Network Profiler, and periodically probing the network.
To this end, each Worker profiler periodically sends a randomly generated file with known file size to each of the other compute nodes via the well-known file transfer protocol called Secure Copy (SCP). 
The file transfer times are recorded and curve-fit using a quadratic regression with respect to the file-size ($f$) as: $l = p + q \cdot f + r \cdot f^2$ where $p,q,r$ are empirically determined constants. 
We opted for a quadratic fit as it is the empirical best fit towards approximate file transfer time for varying fil- sizes.

\subsubsection{Resource Profiler}
The resource profiler provides two types of information to the task mappers: CPU availability and storage availability. 
Each of the Worker Resource Profiler runs a Flask server to listen for resource profiling requests and replies with the current CPU and storage usage upon receiving a request. 
Moreover, each worker profiler periodically sends resource profiling request to every other NCP in the network and stores the collected statistics in a local database.
There is also a provision for immediate polling of resource information where the Worker Profiler returns the current information instead of last known information. 
This is done by instant polling of other NCPs upon receiving such requests from the task mapper or the application. 

\subsubsection{Execution Profiler}
For optimal allocation of tasks, the task mappers such as HEFT might need information about the execution times of the individual tasks for each of the NCPs.
HEFT like task mappers do not use the raw CPU usage statistics available from the Resource Profilers. 
To support such requirements, we have the third and final type of profiler: the Execution Profiler.  
However, the complete execution time information is available only after the tasks are executed on each of the available NCPs. 
To make this information available even before the tasks are actually mapped, the Worker Execution Profiler on each NCP runs the entire DAG with some sample input files and sends the statistics to the Home Execution Profiler. 
Moreover, the Home Execution Profiler also collects information about runtime statistics once the tasks are mapped and executed via CIRCE.

\subsection{Task Mapper}
The task mapper module of the Jupiter is the most important module of the Jupiter system.
As the name suggests, the main function of this module is to optimally map individual tasks of a task-DAG into a set of available compute nodes (NCPs) such that the Makespan of the task DAG is minimized.
To this end, there are two classes of approach that can be opted for: centralized and decentralized. 
In the centralized approach, a central node gathers the global information of the network of compute nodes from the profilers and leverages this information for optimal placement of tasks. 
On the other hand, a distributed approach leverages the local profiling information in each of the compute nodes for task placements. 
In the current version of the Jupiter, we have made available two different classes of task mappers: centralized HEFT and decentralized WAVE.

\subsubsection{HEFT}
Heterogeneous Earliest Finish Time (HEFT)~(\cite{topcuoglu1999task,topcuoglu2002performance}) is a well-known heuristic in grid/cloud computing for mapping a directed acyclic task graph into a network of heterogeneous compute nodes that also accounts for the communication times between the nodes. 
HEFT operates in a sequence of two phases: ranking and prioritization, and processor selection. 
In the first phase, i.e., \emph{ranking or prioritization} phase, HEFT defines a priority of each task $t_i$ as follows:

\vspace{-3ex}
\begin{equation}
    rank_u(t_i) = \overline{\omega_i} + \max_{t_j \in succ(t_i)} (\overline{c_{i.j}} + rank_u(t_j))
\end{equation}
\vspace{-2ex}

\noindent
where the subscript ``u" refers to ``upwards rank" which is defined as the expected distance of the task from the end of the computation, $t_i$ refers to task $i$, $\omega_i$ is the average computation cost of the task $i$ among all the compute nodes, $\overline{c_{i,j}}$ refers to the average communication cost of the data communicated between task $t_i$ and $t_j$ for all pairs of compute nodes, and $succ(t_i)$ refers to the set of dependent tasks in the DAG. For example, the set of dependent tasks for task C in Fig.~\ref{fig:dag}, $succ(C)$, is \{D, E\}.

In the second phase i.e., the \emph{processor selection} phase, HEFT assigns the tasks to the NCPs based on the ranks calculated in the \emph{ranking or prioritization} phase. 
In each iteration of the task assignment, HEFT picks the task which has the highest priority and has all the dependent tasks already mapped. 
Next, HEFT schedules the task on an NCP that will minimize the earliest finish time of that task.
This process continues until all the tasks are mapped.
Finally, HEFT outputs the overall task to NCP mapping along with a timeline to follow for the executions.

\subsubsection{WAVE}
\label{sec:randomwave}

While centralized task mappers are appropriate for cloud computing like scenarios with a network of geographically neighboring compute nodes, a distributed task mapper is more appropriate for Networked Computing due to lower communication and computation overhead as well as fast reaction time.  
To this end, we propose a new class of decentralized task mapper algorithm called the WAVE.
Before detailing WAVE, let us define the notion of \emph{``task controller''} which is an NCP that is in charge of mapping a particular set of tasks from the DAG. 
In the WAVE architecture, there exists a coordinator or home WAVE node (which runs on the Home NCP as illustrated in Fig~\ref{fig:jupiternetwork}) that initiates the whole process, while rest of the nodes, which we refer to as the worker WAVE nodes, perform the actual mapping in a distributed fashion. 
The WAVE algorithm works in two phases as follows.

\begin{figure}[!ht]
    \centering
    \subfloat[]{\label{fig:dag}\includegraphics[width=0.3\linewidth, height=0.5\linewidth]{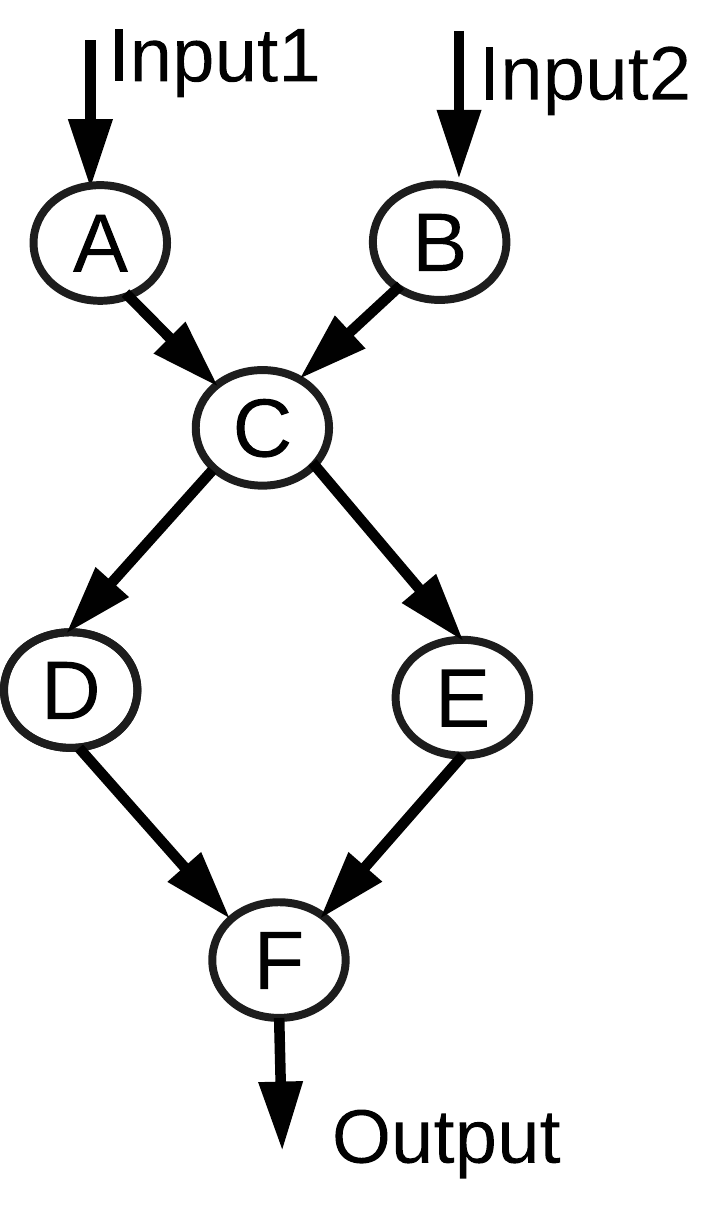}}
    \subfloat[]{\label{fig:wave_il}\includegraphics[width=0.55\linewidth]{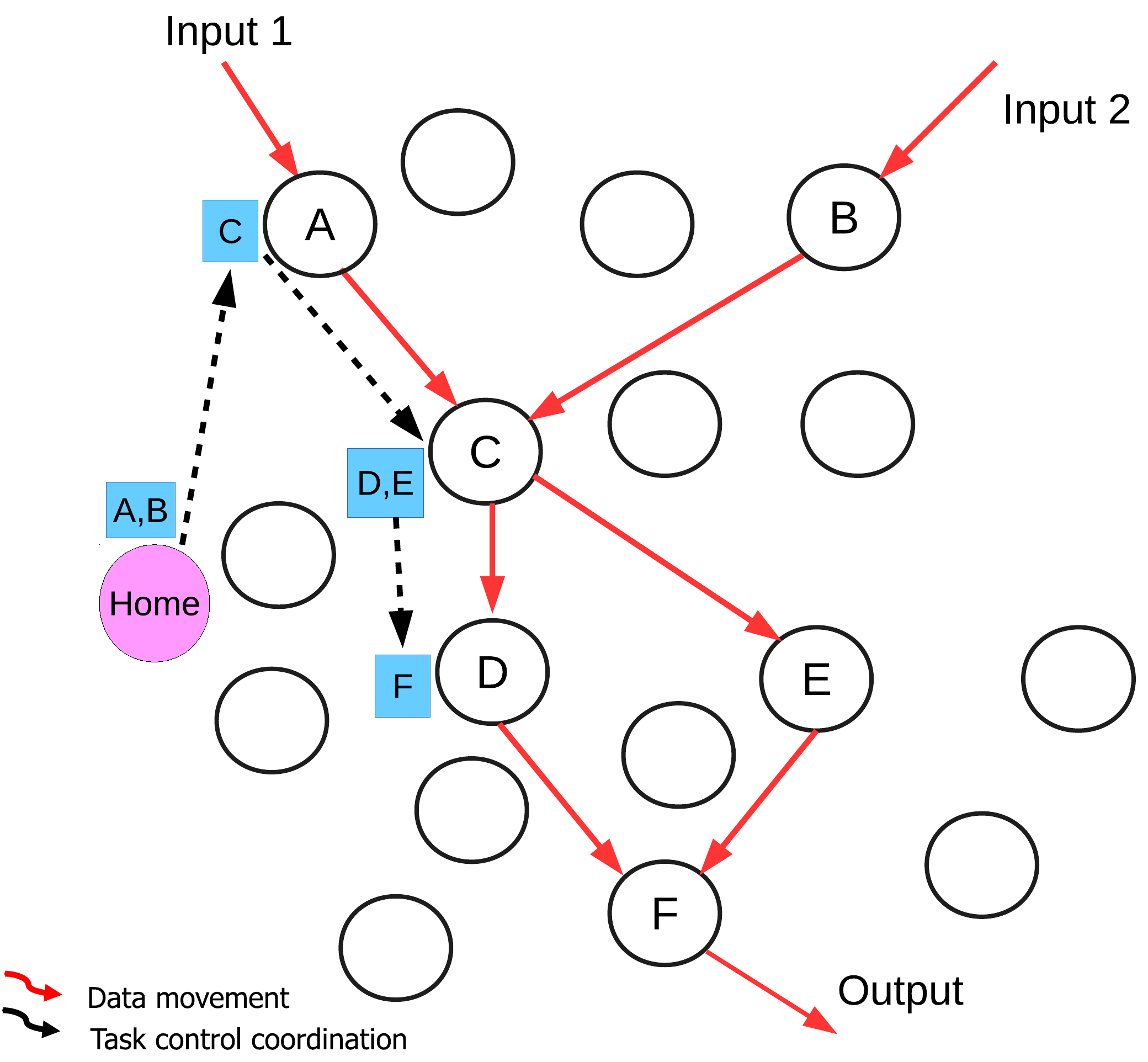}}
    \caption{(a) A Sample Task DAG (b) WAVE Illustration for Task DAG presented in Fig.~\ref{fig:dag}. The square boxes represent the task controllers.}
    \label{fig:wave}
\end{figure}

\textbf{Task Controller Selection:}
In this phase, the WAVE home node chooses a unique ``task controller'' for each task of the DAG. 
For the first level of tasks (e.g., task A and B for the task DAG presented in Fig.~\ref{fig:dag}), the home node itself acts as the task controller. 
For the rest of the tasks, the home node chooses the task controller as follows.
\begin{itemize}
    \item Iterate over the tasks from the DAG in their topological orders. For the sample DAG presented in Fig.~\ref{fig:dag}, one topological order would be \{A, B, C, D, E, F\}.
    \item For each non-input task, check if any of its parent tasks (Tasks A and B are the parent tasks to task C in Fig.~\ref{fig:dag}) are already controllers.
    \item If one of the parents is already a task controller, then appoint that parent as the controller for this task. 
    \item If no parent is already a task controller or multiple parents are task controllers, then choose the parent task with the smaller topological index as the parent.
\end{itemize}
Note that, so far we refer to tasks as task controllers instead of NCPs because at this stage of WAVE the tasks are not mapped/bound to any particular NCP.
In the next step of WAVE, we explain how we map the tasks to the NCPs and consequently map the task controllers to the NCPs.
For illustration purpose, let us assume that the task controller selection output for Fig.~\ref{fig:dag} is $\mathcal{M}_T = \{Home \rightarrow \{A,B\}, A \rightarrow C,  C\rightarrow\{D, E\}, D\rightarrow F \}$ (as shown in Fig.~\ref{fig:wave_il}).
Here $A \rightarrow C$ implies that task $A$ is the task controller for task $C$.

\textbf{Task Mapping:}
In this phase, the WAVE algorithm maps the tasks into appropriate NCPs.
The Home node kick-starts this process by determining the NCPs for each of the input tasks in the DAG, according to the geographical location of the data source.
E.g., for the DAG presented in Fig.~\ref{fig:dag}, the WAVE home will place task A and task B on two NCPs that are near data source 1 and 2, respectively, as presented in Fig.~\ref{fig:wave_il}. 
Note that we assume the data source locations to be known. 
Once the process completes, the WAVE home broadcasts this mapping to the respective NCPs.
Next, the NCPs of the already-mapped tasks will perform similar mapping for the tasks they are in charge of.
For example, the NCP of task A will decide where to run task C which thereafter decides where to run tasks D and E. 
This process continues until all the tasks are mapped. 
Every time a new task mapping is complete, the home node is informed by the respective task controller.
Once the whole process is complete, the WAVE Home returns the mapping information to the next component of Jupiter: the CIRCE dispatcher.
This process is illustrated in Figure~\ref{fig:wave_il}. 

Now, a task controller chooses the optimal NCPs for the tasks by following two different logic as follows.

\textbf{Random WAVE:}

This is the simplest version of WAVE where the task controllers randomly select a NCP from the list of available NCPs.
The task controller does not incorporate the communication and computation costs in the mapping logic which makes the mapping completely non-optimized. This is used as a baseline algorithm and proof of the concept for WAVE.

\textbf{Greedy WAVE:}

The greedy WAVE is a complex version of WAVE that incorporates the profiling information for mapping tasks to the NCPs.
In the greedy WAVE, each of the task controllers connects to the local profilers to get network and resource statistics (it does not use execution profilers). 
Next, each task controller NCP, say NCP $i$, follows a sequence of operation to map the tasks controlled by it.
\begin{itemize}
    \item Based on the end-to-end latency statistics from the network profiler, find the minimum delay $d_{min}^i = \min_{j} d_{i,j} \forall j \neq i \mbox{ and } i,j\in {1,2,\cdots, N}$ where $N$ is the number of NCPs.
    \item Use the calculated $d_{min}^i$ to filter a feasible neighboring compute node set, $S_d^i = \{ j: d_{i,j} < d_{th} \}$ where $d_{th} = k \cdot d_{min}^i$ is the threshold latency. We have empirically chosen an value of $k=15$ for our experiments due to a very wide distribution of network delay. 
    \item Use the resource information i.e., the CPU usage, $p_j$, and the memory usage $m_j$ of each neighbor $j$ to rank the neighbors in $S_d^i$ as follows:
    \begin{equation}
        rank(i,j) = \omega_d \cdot d_{i,j} + \omega_p \cdot p_j + \omega_m \cdot m_j \ \forall j \in S_d^i
        \label{eqn:wavegreedy}
    \end{equation}
    where $\omega_d$, $\omega_p$, $\omega_m$ are three weighing constants to determine the rank. We empirically choose a value of $\omega_d= \omega_p= \omega_m = 1/3$ for the experiments presented in this paper.
    \item Use the rank information to map the tasks. If the task controller is responsible for $n$ tasks and $n\leq |S_d^i|$, it maps the tasks to $n$ top ranked neighbors based on the ordering of the tasks on the DAG. Otherwise, first map the top $|S_d^i|$ tasks on the $|S_d^i|$ neighboring NCPs (one-to-one) according to the rank and task ordering. This is followed by repeating the same process for rest of the tasks. 
\end{itemize}

\subsection{The CIRCE Dispatcher}
\label{sec:circe}
The CIRCE dispatcher is the third and the final module of the Jupiter system.
The CIRCE dispatcher is the part of Jupiter that inputs the task-to-NCP mapping and dispatches the tasks on the respective nodes.
CIRCE wraps the task codes to support an input-output queuing system.
CIRCE creates an input folder/queue and an output folder/queue for each task and takes care of transferring the output of a task to the input the next task in the DAG using the well-known SCP tool.
Every time a new data file arrives at the input folder, CIRCE starts the execution of the respective task. 
Sometimes, a task might require the output of more than one parent task as its input. 
In such cases, CIRCE also takes care of waiting for all the inputs to arrive before starting the execution.
At the completion of execution, CIRCE starts the file transfer process to the next task.
If there are multiple child tasks, CIRCE transfers a copy of the output to the input of each of the child tasks.
CIRCE uses a sequence number for the ordering of the input data.


\section{Implementation Details}
Jupiter is implemented on top of a well-known open-source cloud-based-container-orchestrator from Google called the Kubernetes (\cite{bernstein2014containers,hightower2017kubernetes}). 
Before explaining the implementation, let us briefly introduce some key concepts such as Container, Docker, and the Kubernetes architecture.

\subsection{Containers, Docker, and Kubernetes}
The most common norm in cloud computing today is to use Virtual Machines (VMs) to support the demand of users while keeping necessary isolation between the user processes running on the same physical machine.
VMs run inside a guest OS and accesses the guest Hardware via the concept of virtual Hardware. 
To support this sort of isolation from the real hardware as well as among different VMs, the computational overhead of VM is substantial.
This limits the number of concurrent VMs on a Physical machine to a very small number such as 3-5 for a standard Quadcore desktop with 12GB RAM. 
Moreover, due to the lack of direct access to the Hardware, the functionality of VMs are restricted.
Containers, on the other hand, are the most cutting-edge convention for processor virtualization that provides isolations similar to traditional VMs but with much less computing power requirements.  
Unlike VMs, a container image is a lightweight standalone executable that includes all the requires modules, libraries, codes, and tools to run it.
A container directly runs on the guest OS and is considered as a single process by the guest OS. 
All the processes inside a container are viewed as a sub-process of the main process. 
Because of this low computation requirement, one can run hundreds of containers on a physical machine.
The concept of a container has been around for a while but was not popular until the advent of a specific type of containers called the Dockers~\cite{merkel2014docker} in mid-2014. 
Among the handful of available container orchestration tools, Google Kubernetes, Apache Mesos, and Docker Swarm are the most promising ones.
Out of these options, we opted for Kubernetes due to its popularity as well as its unique features such as a support of Raspberry Pi3 devices. 
Briefly speaking, in Kuberenetes, there exists one central high power master node (which we refer to as the K8 Master) that maintains a network of compute nodes, keeps tracks of the deployed containers, and restarts the deployed containers in case of failures.
Note that the K8 Master is different from the Home NCP required for Jupiter system which can be run on the same NCP or different NCPs. 
A detailed overview of Kuberenetes can be found in the official website \url{https://kubernetes.io/}{https://kubernetes.io/}.

\subsection{Jupiter on Kubernetes}
We implemented each component of the Jupiter in Dockers to support parallelism, have isolation between different Jupiter modules, support scalable and easy administrations, and support multiple simultaneous DAGs.
Moreover, by using containers, every module of Jupiter is uniquely addressable (via unique IP and port numbers) which helps in a uniform system implementation.  
Another reason behind the Dockerization of Jupiter is to make it compatible with the Kubernetes system.
Next, we briefly detail different types of Dockers used in the Jupiter system. 

\textbf{Network and Resource Profiler Dockers:} 
For compactness, we put the network and resource profiler inside one Docker instead of separate ones.
We combine network home profiler and resource home profiler into a combined Docker which runs on the Home NCP.
Similarly, we combine the worker network profiler and worker resource profiler into a single Docker which runs on each NCP of the network except the Home NCP. 

\textbf{Execution Profiler Dockers:}
Due to different functionality than the network and resource profilers, we have kept the execution profilers in a separate Docker.
Again, we create two different types of execution profiler Docker to correspond to the home execution profiler (runs on the Home NCP) and the worker execution profiler (runs on all NCPs expect the Home NCP), respectively.

\textbf{HEFT Profiler Docker:}
Because HEFT is a centralized profiler, there is only one Docker needed for HEFT which can run on any NCP of the network. However, for consistency, we choose to run it on the Home NCP.

\textbf{WAVE Dockers:}
For both WAVE Greedy and WAVE Random, we have the notion of home and worker. Therefore, we need two separate Dockers for ``WAVE-home'' and ``WAVE-worker'' that runs on the Home NCP and the rest of the NCPs, respectively. 

\textbf{CIRCE Dockers:}
The experimental implementation of CIRCE has the notion of home and worker as well. 
Here, CIRCE home is used mainly to emulate a data source as well as to collect different statistics whereas a CIRCE worker Docker follows the description presented in Section~\ref{sec:circe}. Therefore, we have two Dockers for CIRCE as well: ``CIRCE-home'' and ``CIRCE-worker''. The CIRCE worker Docker contains all the task files but can run only one task of the DAG at a time. Thus, the number of CIRCE worker Dockers on the network equals the number of tasks in the task-DAG. If an NCP has multiple tasks allocated to it, it will run multiple CIRCE Dockers. 

    
    
    
    


\section{Experimental Results and Analysis}
\begin{figure}[!ht]
    \centering
    \subfloat[]{\label{fig:geo}\includegraphics[width=0.4\linewidth]{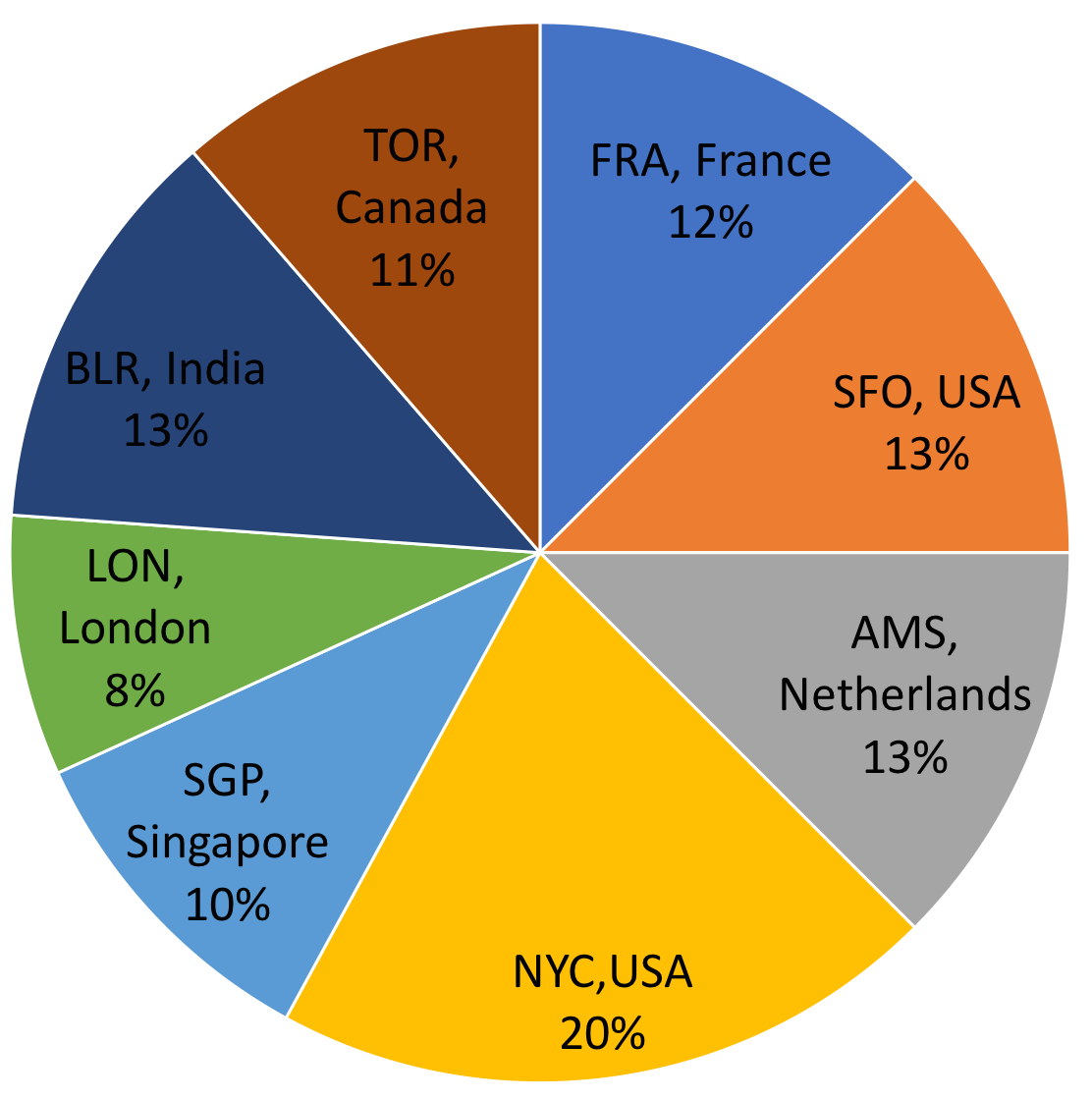}}\quad
    \subfloat[]{\label{fig:rpi}\includegraphics[width=0.53\linewidth]{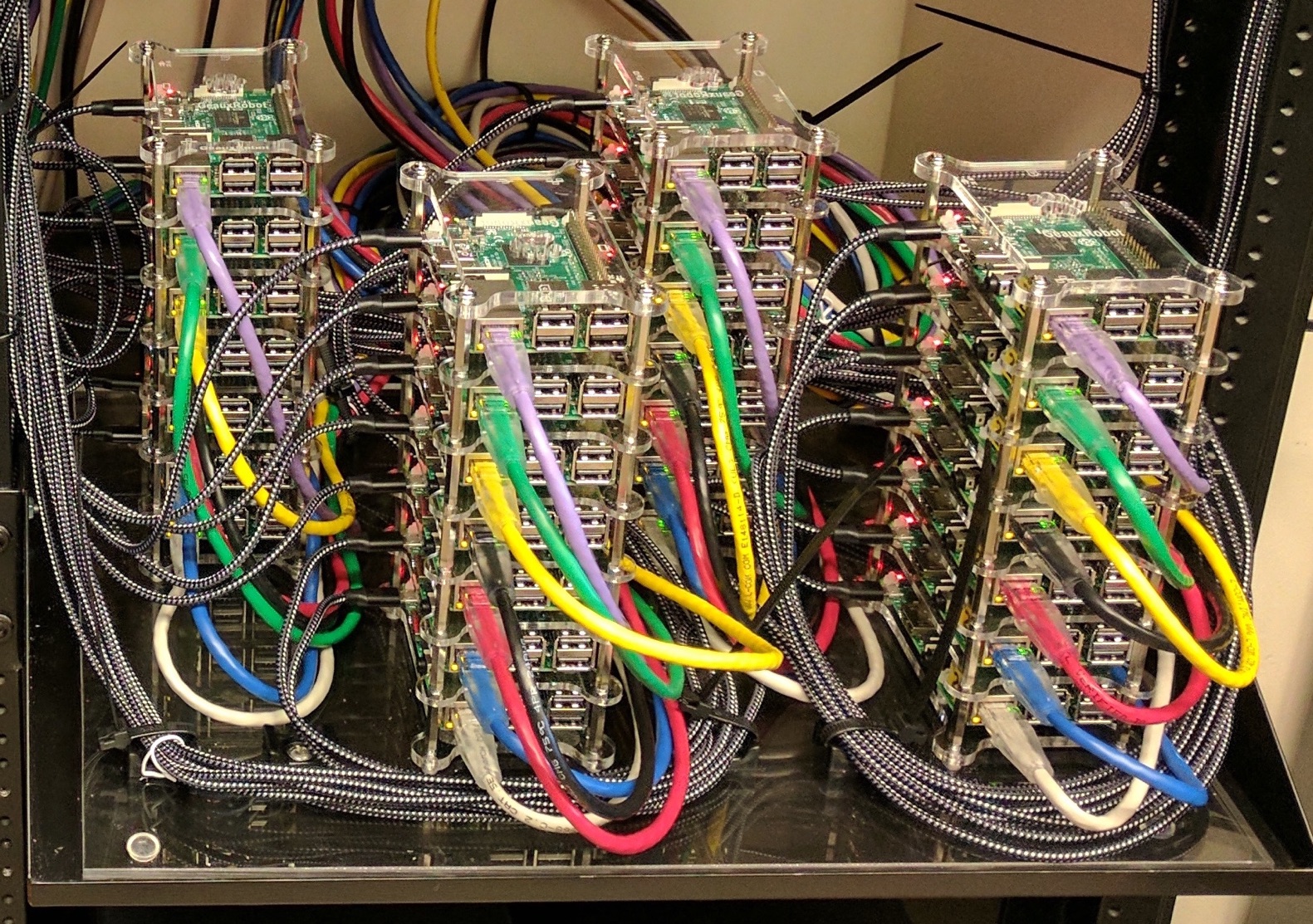}}
    \caption{(a) Geographical Distribution of the Digital Ocean VMs and (b) the RPI Cluster}
\end{figure}

We analyze the performance of the Jupiter system via a range of experiments with the DNAD application.
For these experiments, we use two different clusters with Kubernetes.
The first cluster consists of 90 Virtual machines, also referred to as Droplets, from a cloud provider called the Digital Ocean. 
Out of the 90 VMs, 13 VMs have 2GB RAM while the rest have 3GB RAM.
Each of these VMs has 20 GB of disk space available.
We handpicked the set of VMs from 8 available geographic locations across the world.
The geographic distribution of nodes is presented in Fig.~\ref{fig:geo}. 
We use an 8GB VM as the Kubernetes master node for this cluster and for the Home NCP, we randomly pick one from the 90 VMs.



The second testbed consists of 30 in-house Raspberry Pi3 nodes with 1GB RAM and 64GB SD cards, as illustrated in Figure~\ref{fig:rpi}. 
They are connected via a Cisco switch to control the network conditions and topology.
For the Kubernetes cluster implementation, we have opted for a mixed architecture where an AMD64 Virtual Machine with 8GB RAM and 10GB disk space acts as the Kubernetes master while all the actual compute nodes are ARM32 real processors on the Raspberry Pis. Again, the Home node is chosen at random from the 30 RPIs. 

In both testbeds, we ran the Jupiter system with all three types of task mappers i.e., HEFT, 
WAVE Random, and WAVE Greedy.
For each configuration of Jupiter, we ran the whole Jupiter system 25 times. 
In each run, once the CIRCE deployment is complete, we feed a sequence of 10 pre-loaded files with file sizes in the range of 10KB to 300KB and record different statistics such as the Makespan of the task DAG and the execution times of individual tasks. 
Note that, we refer to HEFT as original HEFT in this section as we propose some modification to the HEFT based on the experiment results to improve its performance.

\subsection{DAG Makespan Analysis}

\begin{figure}[!ht]
    \centering
    \subfloat[]{\label{fig:rpi-Makespan}\includegraphics[width=\linewidth]{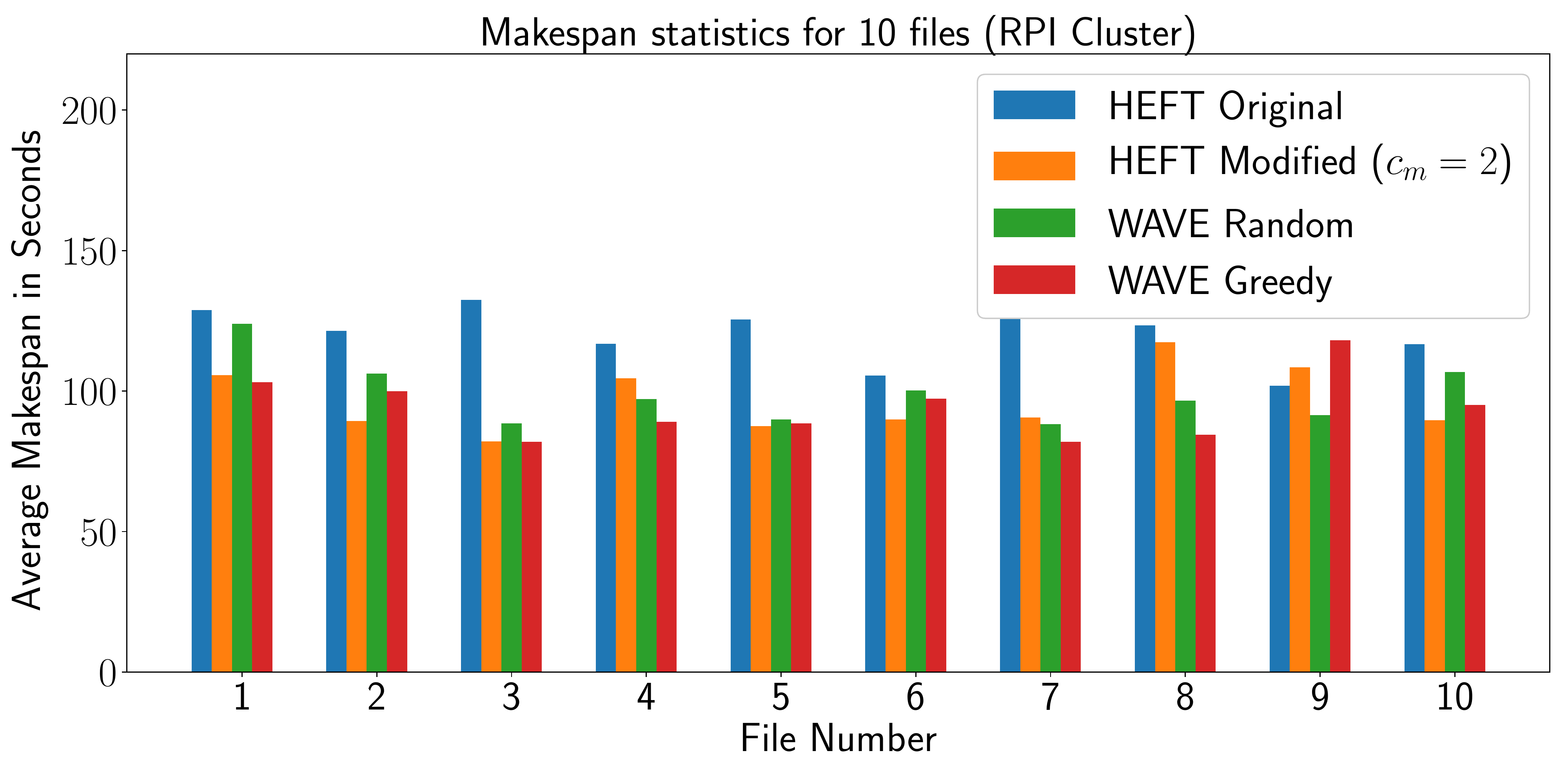}}\,
    \subfloat[]{\label{fig:do-Makespan}\includegraphics[width=\linewidth]{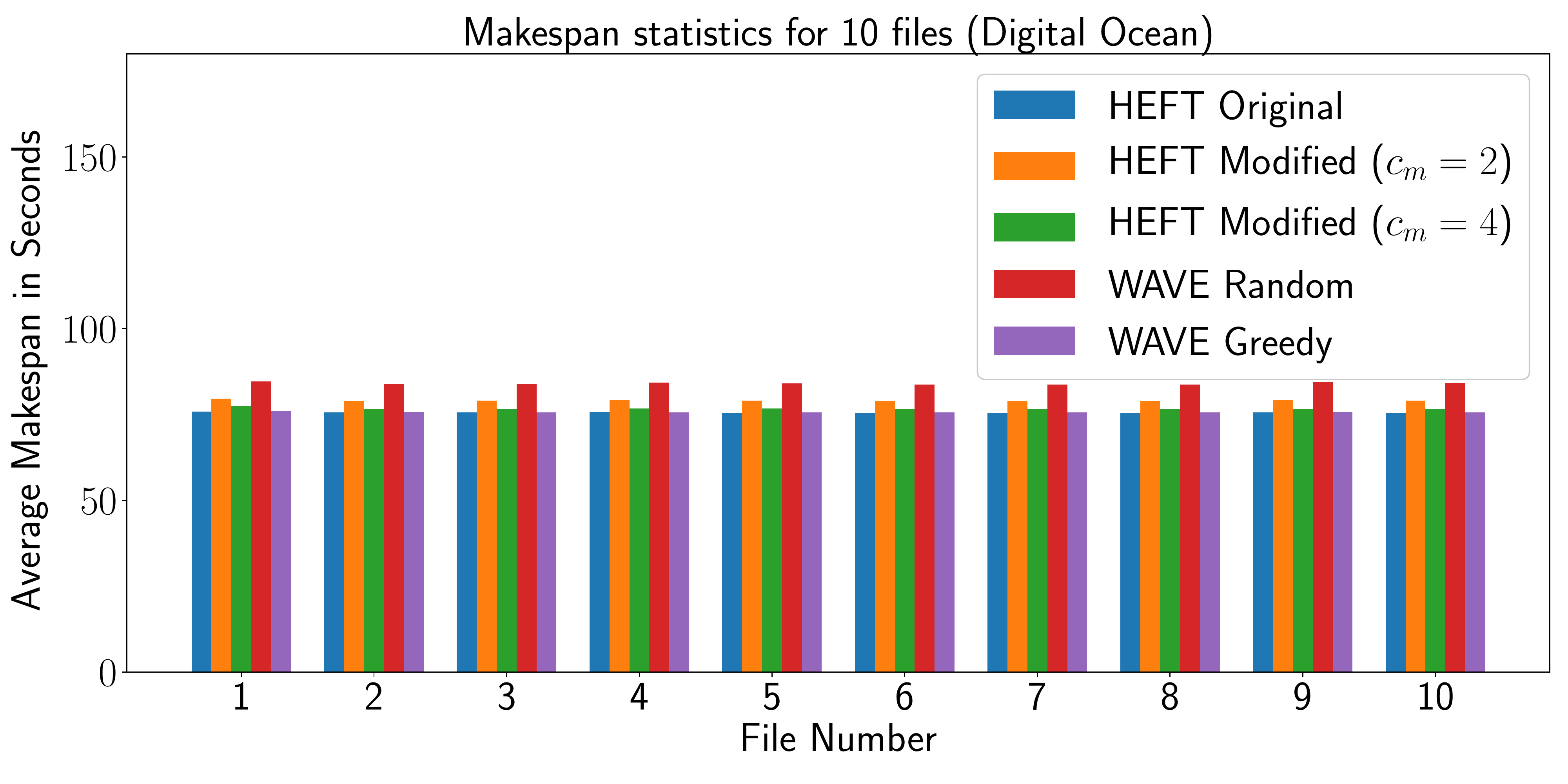}}
    \caption{Makespan Statistics of the DNAD Task-DAG for 10 Different Files in (a) the RPI Cluster and (b) the Digital Ocean Cluster.}
    \label{fig:makespn}
\end{figure}

In the first set of experiments, we compare the Makespan statistics of the DNAD application for different task mapping algorithms. 
In Fig.~\ref{fig:rpi-Makespan}, we present the Makespan statistics for the RPI cluster for 10 sequential inputs files which shows that among the original HEFT, WAVE Greedy, and WAVE Random algorithm, the performance of the WAVE Greedy is the best, followed by the performance of WAVE random and HEFT.
While the bad performance of the WAVE random is precedented due to the random nature of the mapping algorithm, the bad performance of the original HEFT was not warranted since HEFT is a well-respected heuristic for DAG based task graph in cloud/grid computing.
Upon further investigations, we find that this bad performance of HEFT is rooted at the processing power and stability limitations of the RPIs. 
We find that resource-constrained devices like RPI3 are very unstable and failure prone if we run too many CIRCE containers (more than 3-4 CIRCE containers) alongside with the other required components (i.e, the profilers and task mappers) of the Jupiter. 
The original HEFT works under the assumption that the task executions will follow the timeline suggested by the algorithm.
However, to support the continuous execution of incoming files and parallel processing in the NCPs, Jupiter keeps a separate CIRCE docker running for each of the tasks.
When HEFT tries to optimize the Makespan by reducing communication overhead and putting many tasks on the same NCP, it ends up overloading the RPIs. 
While the Jupiter system can recover from failures, multiple failures of the overloaded RPIs actually ends up adding more delay in the execution of the tasks as well as the communication between tasks due to temporary disruptions of the data flow. 
To circumvent this issue, we propose a minor modification to the original HEFT where HEFT is restricted to allocate no more than $c_{m}$ containers per NCP where the number $c_{m}$ is dependent upon the processing power of the node.
We empirically choose a value of  $c_{m}=2$ for the RPI3 cluster. 
We will refer to this version of HEFT as the modified HEFT.
The performance analysis of the modified HEFT (presented in Fig.~\ref{fig:rpi-Makespan}) shows that, by this slight modification, the performance of HEFT is improved and becomes comparable to the WAVE Greedy. 

We perform a similar set of experiments on the Digital Ocean cluster and present the results in Fig.~\ref{fig:do-Makespan}.
Figure~\ref{fig:do-Makespan} shows that the performance of HEFT Original and the WAVE Greedy are comparable while the performance of the WAVE Random is the worst followed by the modified HEFT. 
Again the bad performance of WAVE Random is due to random selection of the NCPs without really accounting for any communication and processing overheads.
The reason behind modified HEFT not performing well is that a stable cloud system cluster like the Digital Ocean with higher computing power can accommodate more than 2 CIRCE dockers per NCP. 
By putting the restriction on HEFT, we are forcing HEFT to choose a different NCP which thereby adds networking delay in the Makespan calculation.
To verify this, we perform another set of experiments on the Digital Ocean cluster with modified HEFT and $c_{m}= 4$. 
The results presented in Fig.~\ref{fig:do-Makespan} shows that with $c_m =4$ the performance of modified HEFT is similar to original HEFT. 
This suggests that the modified HEFT have similar or better performance than the original HEFT provided that $c_m$ properly account for the processor limitations. 

In summary, modified HEFT with a properly selected value of $c_m$ and WAVE Greedy have the best performance among all four choices of mappers (Original HEFT, Modified HEFT, WAVE Greedy, and WAVE Random) in Jupiter. 
The results also substantiate that a distributed algorithm which relies only on local information can have similar performance as a centralized globally-informed algorithm.

\subsection{Runtime Analysis of the Task Mapping Algorithms}
We perform another set of experiments to analyze the runtime of the modified HEFT, Greedy WAVE, and the Random WAVE algorithms.
For these experiments, we do not consider the original HEFT as the runtime of original HEFT is similar to the runtime of the modified HEFT.
The runtime statistics are presented on Table~\ref{tab:runtime}.
Table~\ref{tab:runtime} shows that for the 30 node RPI cluster, the runtime of HEFT is the lowest and the runtime of WAVE Random and WAVE Greedy are similar.
Ideally, the HEFT algorithm runtime should be larger than WAVE as HEFT requires all the profiler data to be gathered at a central point which takes a considerable amount of time whereas WAVE relies only on local information. 
Moreover, the amount of data to be processed in HEFT is larger than the amount of data to processed by each of the NCPs in WAVE. 
As a reason behind this counter-intuitive result, we hypothesize that for a cluster of that size (30 nodes) the communication time overhead for sequential task assignments via task controllers over the network (as in WAVE) is much larger than the time required to gather all the statistics in a central location (as in HEFT).
To verify this hypothesis, we run a similar set of experiments on the Digital Ocean Cluster with 30 nodes, 60 nodes, and 90 nodes. 
The results presented in Table~\ref{tab:runtime} shows that as the number of nodes increases, the runtime of HEFT gradually becomes comparable (for 60 nodes) and even worse than Greedy WAVE (for 90 nodes).
This suggests that with increasing network size, the time required to gather all the statistics at a central location increases and eventually surpasses the communication time overhead for sequential task assignments in WAVE. 

Another interesting observation is that WAVE Random runtime is actually slightly larger than the WAVE Greedy whereas intuitively the runtime of WAVE random should be smaller.
Upon further investigation, we discovered that this inconsistency between the expected result and the observed result is due to the communication delay difference between the task controllers. 
By randomly assigning the NCPs, the end-to-end delay in communication between the task controllers, which is required for the mapping purpose, becomes larger compared to the same for WAVE Greedy.
We also discover that this delay is the largest component of the runtime of WAVE.
For illustration, in the 90 node cluster, the time required to retrieve the network data is $\approx 0.05s$ and the resource data is $\approx 10s$ whereas the runtime of the WAVE Greedy is around $100s$. 
This suggests that there is still a lot of room for improvement in the WAVE algorithm.

Lastly, it is evident from Table~\ref{tab:runtime} that the runtime of the WAVE algorithm remains almost same regardless of the cluster size as it mainly depends on the number of tasks in the DAG rather than the actual number of NCPs.
On the other hand, the performance of the HEFT algorithm varies proportional to the number of NCPs in the network.
This makes WAVE much more scalable than HEFT for a distributed networked computing system.

\begin{table}[!ht]
    \centering
    \caption{Runtime Statistics of the Mapping Algorithms in Seconds}

    \begin{tabular}{|p{1.6cm}|c|c|c|c|c|c|}
    \hline
     \multirow{2}{1.5cm}{Cluster Details}  &  \multicolumn{2}{c|}{Modified HEFT} & \multicolumn{2}{c|}{Random WAVE} & \multicolumn{2}{c|}{Greedy WAVE} \\ \cline{2-7}
       & Mean  & STD  & Mean  & STD & Mean  & STD \\ \hline
    {RPI Cluster (30 NCPs)}  & 102.48 & 69.27  & 156.37 & 78.74   &  154.94 & 73.80 \\ \hline
    
    {Digital Ocean (30 NCPs)} &  55.73 & 25.73   & 90.48 &  1.48   & 89.94 & 0.92  \\ \hline 
    
    {Digital Ocean (60 NCPs)} & 89.39 & 32.32 & 111.06  &  64.62   & 92.40 & 8.12  \\ \hline 
    {Digital Ocean (90 NCPs)}  & 136.36  & 45.52   & 101.35  &  11.45   & 97.44 & 2.52  \\ \hline 
    
    \end{tabular}
    \label{tab:runtime}
\end{table}

\section{Related works}
\label{sec:related}

The wide range of related works mainly lies within the rich literature of cloud computing, edge computing, and grid computing. 

The well-known field of cloud computing deals with virtualization technology that enables data centers to provide a range of services such as computing service and storage service to the users on pay-per-use basis~\cite{Singh2016}. 
The typical cloud computing architecture with hundreds of powerful co-located servers requires all the data to be first collected at the cloud before the actual computation can be performed~\cite{dinh2013survey}.  
In contrast, the Jupiter system enables geographical proximity based mapping of the tasks without requiring the data to be collected at a central point. 

There also exists some challenges of cloud computing such as the data pre-processing from heterogeneous unstructured data sources, intermittent connectivity of the data sources, and low latency requirement~\cite{HASHEM201598} that can be solved both by edge computing tools~\cite{Mach2017} and networked computing tools (e.g., Jupiter). 
However, edge computing architectures has its own set of limitations such as limited geographical span and limited amount of resources~\cite{Mao2017MobileEC}.  
In contrast, our proposed Jupiter system considers a geographically dispersed set of heterogeneous compute nodes (they can be both cloud nodes or edge nodes) and distributes the tasks among them while minimizing the Makespan, if the application is represented as a DAG. 

In the grid computing domain~\cite{berman2003grid}, there also exists some frameworks for mapping tasks on geographically distributed clusters such as Pegasus~\cite{deelman2005pegasus} and Falcon~\cite{raicu2007falkon}.
However, they assume a static and relatively well characterized network with simple applications for centralized task mapping.
In comparison, the Jupiter framework along with the WAVE scheduling algorithm is designed to support complete geographical separation along with decentralized task mapping for any complex DAG based applications.
Moreover, Jupiter can be easily modified into a cloud computing, edge computing, or even a grid computing system, if required.


In the context of cloud computing, there exists a range of scheduling algorithms that schedules tasks from a DAG into multiple cores of a single or multiple co-located processors~(\cite{Guo2017,Keshanchi2017}) without accounting any communication costs. 
On the other hand, there exist some centralized schedulers for DAG-based applications that do account for the communication overhead between multiple processors such as the Heterogeneous Earliest Finish Time (HEFT)~\cite{topcuoglu2002performance}, the Longest Dynamic Critical Path (LDCP) algorithm~\cite{Daoud2008}, the Dynamic Level Scheduling (DLS)~\cite{sih1993compile}, and the Critical Path On a Cluster (CPOC) algorithm~\cite{topcuoglu2002performance}. 
Nonetheless, these algorithms are mainly aimed at geographically co-located processors with considerably low and static delay statistics. 
Our proposed WAVE algorithm, on the other hand, is designed mainly for a sparse and dynamic network of compute nodes. 

While, there also exists some state-of-the-art algorithms for distributed scheduling~(\cite{huang2013exploring,hamscher2000evaluation,ranjan2008decentralized}), to our knowledge, only a few of them are directly applicable to DAG based task scheduling on a networked computing cluster. 
Nonetheless, the Jupiter system can support a suitably enhanced version of any of these scheduling algorithms, if needed. 



\section{Conclusion}
In this paper, we proposed a new distributed system for Networked Computing called the Jupiter that can efficiently distribute and deploy tasks from a DAG based task graph with the goal of Makespan minimization.
We also proposed a new class of distributed task-mapping or scheduling algorithms called the WAVE that leverages local information of the NCPs for distributed mapping of the tasks.
Through a wide range of real-world experiments on a 90-node (distributed across 8 cities in 7 countries) Cloud-based cluster and a 30-node edge computing cluster, we show that WAVE can perform similar or even better than a globally informed centralized task-mapping algorithm called the HEFT. 
However, while WAVE shows promising attributes in terms of Makespan performance and scalability, it lacks any optimality guarantee and there is still a lot of room for improvements. 
Therefore, in our future works, we would like to explore for an optimal settings of WAVE.
We would also like to explore more centralized and decentralized mapping algorithms to find a better and more generic trade-off between centralized and decentralized algorithms for Networked Computing. 
Lastly, we would like to introduce the concept of mapping-algorithm-independent load balancing in Jupiter to support dynamics in incoming data traffic without having to re-run the task Mapping algorithm. 

{
\bibliographystyle{IEEEtran}
\bibliography{IEEEabrv,ref}
}
\end{document}